\documentstyle[11pt,graphicx]{article}
\input epsf 

\begin{document} 

%
\newcommand\rd{{\rm d}}
\newcommand\p{\partial}
\newcommand\nab{\nabla}
\newcommand\pr{\prime}
\newcommand\q{\quad}
\newcommand\qq{\qquad}
\newcommand\r{\right}
\renewcommand\l{\left}

\newcommand\e{{\epsilon}}
\newcommand\ga{{\gamma}}
\newcommand\th{{\vartheta}}

\newcommand\bA{{\bf A}}
\newcommand\bB{{\bf B}}
\newcommand\bF{{\bf F}}
\newcommand\bJ{{\bf J}}
\newcommand\bX{{\bf X}}
\newcommand\bu{{\bf u}}
\newcommand\bx{{\bf x}}
\newcommand\bnu{{\mbox{\boldmath $\hat\nu$}}}
\newcommand\bom{{\mbox{\boldmath $\omega$}}}
\newcommand\bxi{{\mbox{\boldmath $\xi$}}}
\newcommand\Mu{{M}}
\newcommand\Nu{{N}}

\newcommand\cC{{\cal C}}
\newcommand\cD{{\cal D}}
\newcommand\cF{{\cal F}}
\newcommand\cN{{\cal N}}
\newcommand\cR{{\cal R}}
\newcommand\cS{{\cal S}}
\newcommand\cT{{\cal T}}

\newcommand\bb{{\hat{\bf b}}}
\newcommand\bn{{\hat{\bf n}}}
\newcommand\bt{{\hat{\bf t}}}
\newcommand\beo{{{\hat{\bf e}}_1}}
\newcommand\bet{{{\hat{\bf e}}_2}}
\newcommand\ber{{{\hat{\bf e}}_r}}
\newcommand\beth{{{\hat{\bf e}}_\vartheta}}
\newcommand\bethr{{{\hat{\bf e}}_{\vartheta_{R}}}}
\newcommand\fc{{f_c}}
\newcommand\ft{{f_\tau}}
\newcommand\s{{\hat{s}}}
\newcommand\A{{\tilde{A}}}
\newcommand\B{{\tilde{B}}}
\newcommand\C{{\tilde{C}}}
\newcommand\D{{\tilde{D}}}
\newcommand\E{{\tilde{E}}}
\newcommand\F{{\tilde{F}}}
\newcommand\G{{\tilde{G}}}
\renewcommand\H{{\tilde{H}}}
\newcommand\I{{\tilde{I}}}
\newcommand\J{{\tilde{J}}}
\newcommand\Bs{{B_s}}
\newcommand\bBa{{{\bf B}_a}}
\newcommand\bBm{{{\bf B}_m}}
\newcommand\Bth{{B_\theta}}
\newcommand\thr{{\vartheta_R}}
\newcommand\bTn{{{\hat{\bf T}}_2}}
\newcommand\bTb{{{\hat{\bf T}}_3}}

\newcommand\NN{{\hbox{I\kern-.14em{N}}}}
\newcommand\RR{{\hbox{I\kern-.14em{R}}}}
\newcommand\ZZ{{\hbox{I\kern-.14em{Z}}}}

\newcommand{\qed}{{\par\hfill $\Box$\medskip}}


\pagestyle{empty}
\parindent=0mm
\null\vspace{-20mm}
\begin{center}
{\LARGE\bf On the groundstate energy of tight knots}\\ 
\vspace{10mm}
\large{\textsc{Francesca Maggioni}${\;}^{1}$ and 
\textsc{Renzo L. Ricca}${\;}^{2,\!}$
\footnote{Corresponding author: \tt renzo.ricca@unimib.it}}\\
\vspace{3mm}
{\it ${}^{1}$ Dept. Mathematics, Statistics, 
Computer Science \& Applications, U. Bergamo}\\
{\it Via dei Caniana 2, 24127 Bergamo, ITALY}\\
and\\
{\it ${}^{2}$ Dept. Mathematics \& Applications, U. Milano-Bicocca}\\
{\it Via Cozzi 53, 20125 Milano, ITALY}\\ 
\vspace{5mm}
(Revised: 2 April, 2009)\\
\vspace{10mm}
ABSTRACT\\
\end{center}
New results on the groundstate energy of tight, magnetic knots are
presented.  Magnetic knots are defined as tubular embeddings of the
magnetic field in an ideal, perfectly conducting, incompressible
fluid.  An orthogonal, curvilinear coordinate system is introduced and
the magnetic energy is determined by the poloidal and toroidal
components of the magnetic field.  Standard minimization of the
magnetic energy is carried out under the usual assumptions of volume-
and flux-preserving flow, with the additional constraints that the
tube cross-section remains circular and that the knot length 
(ropelength) is independent from internal field twist (framing). 
Under these constraints the minimum energy is determined analytically 
by a new, exact expression, function of ropelength and framing.
Groundstate energy levels of tight knots are determined from
ropelength data obtained by the SONO tightening algorithm developed by
Pieranski (Pieranski, 1998) and collaborators.  Results for torus
knots are compared with previous work done by Chui \& Moffatt (1995),
and the groundstate energy spectrum of the first prime knots --- up
to 10 crossings --- is presented and analyzed in detail. These results
demonstrate that ropelength and framing determine the 
spectrum of magnetic knots in tight configuration.
\vfill
\begin{flushleft}
{\bf Short title:} Groundstate energy of tight knots
{\bf Submitted to:} {\em Proc. R. Soc. \rm A} 
\end{flushleft}
\eject
%
%
\pagestyle{myheadings}
\parindent=10mm

\section{Introduction}
Work to establish rigorous relationships between energy and
topological complexity of physical systems is of fundamental
importance in both pure and applied mathematics.  Progress in this
direction has been slow, but steady, since Arnold's original
contribution of 1974 (Arnold, 1974).  For magnetic knots, in
particular, the problem can be synthesized as follows: if the initial
field is confined to a single knotted flux tube of signature
$\{V,\Phi\}$ ($V$ magnetic volume and $\Phi$ magnetic flux), then the
minimal magnetic energy $M_{\rm min}$ under a signature-preserving
flow is given by (Moffatt, 1990)
\begin{equation}
    M_{\rm min}=m(h)\Phi^{2}V^{-1/3}\ ,
    \label{moffatt}
\end{equation}
where $m(h)$ is a positive dimensionless function of the 
dimensionless twist parameter $h$. Of particular interest is the 
value of $h$ for which $m(h)$ is minimal ($m_{\rm min}$). A 
fundamental problem here is this (Moffatt, 2001):

\vskip4pt plus2pt
\noindent
{\bf Problem 1.} {\em Determine $m_{\rm min}$ for knots of minimum crossing 
number 3, 4, 5, \ldots .}     
\vskip4pt plus2pt

Minimization of magnetic energy of knot types bears some analogies
with another type of problem, that originates from work on the shape
of ideal knots (see, for instance, the collection of papers edited by
Stasiak \emph{et al.}, 1998).  In this context knots are thought of as
made by a perfectly flexible and infinitely hard, cylindrical rope
closed upon itself; a fundamental question here is given by the
following (Litherland \emph{et al.}, 1999):

\vskip4pt plus2pt
\noindent
{\bf Problem 2.} {\em Can you tie a knot in a one-foot length of one-inch 
rope?}     
\vskip4pt plus2pt

This problem admits an obvious generalization to knot types of
increasing complexity, so that the question can be generalized as to
finding the minimal length of a given knot type.  As we shall see, the
two problems tend to coincide at some fundamental level.  If the
relaxation of magnetic field to minimum energy state occurs under a
volume- and flux-preserving flow, then the process, driven by the
action of the Lorentz force, resembles the minimal shortening of an
infinitely flexible, elastic rope under shrinking deformation.  In the
incompressible limit, a shrinking, volume-preserving flow acts on the
tubular knot by increasing the average tube cross-section as the knot
length diminishes.  Thickening of the fattening knot stops when the
topological barriers given by the knot type prevents further relaxation 
(see \S \ref{constrained} below). For magnetic knots this end-state will 
have minimal energy \emph{and}, for tight knotted ropes, minimal ropelength.
Existence of a positive lower bound for magnetic energy (Freedman, 1988), 
however, is not sufficient to guarantee that a global minimum is actually 
attained, even in ideal conditions. With increasing knot complexity, for 
example, configurational arrangements may indeed prevent full 
relaxation, with local minima of magnetic energy (or ropelength)  
trapped from further minimization. 

In the present paper we shall consider magnetic relaxation subject to
the invariance of magnetic signature (volume and flux) with the
\emph{a priori} assumption that the magnetic tubular boundary of the
flux tube remains circular at all times and that the knot length is
independent from the internal twist $h$.  There is certainly no
physical reason to expect this to happen, other than mathematical
advantage in the analysis of the energy functional.  These assumptions
pose additional mathematical constraints on the relaxation process,
preventing full minimization.  However, the information on energy thus
found provides, we believe, a reasonable approximation (from above) to
the true bound.  By this approach we shall demonstrate (\S
\ref{constrelaxation}) that magnetic energy minima can be related to
the minimal ropelength by an exact, analytical expression (given by
eq.  \ref{Mresult}, or the simplified form \ref{mr}) for the minimized
constrained magnetic energy of knots.  Then, by using minimal length
data obtained by the SONO algorithm (briefly reviewed in \S
\ref{SONO}) developed by Pieranski (1998) and collaborators, we
determine the constrained groundstate energy levels of the first 250
prime knots up to 10 crossings (\S \ref{last}).  In doing so, we also
compare results extrapolated from the SONO data by using eq.
(\ref{mr}) with previous work done by Chui \& Moffatt (1995; hereafter
denoted by CM95), highlighting the limitations of their approach and
commenting on some marked differences in the results.  Some critical
issues and open problems for future work are discussed in the final
section \S \ref{least}.

\section{Magnetic knots as tubular embeddings in ideal fluid}
\begin{figure}[t]
\begin{center}  
\epsfysize=60mm\epsfbox{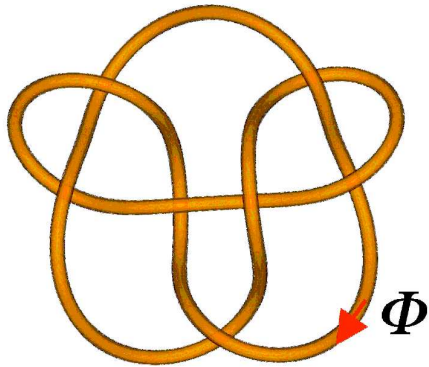}
\caption{Magnetic knot given by the tubular embedding 
of the magnetic field centred on the knot $K_{7.7}$. 
In ideal conditions knot volume $V$ and magnetic flux $\Phi$ 
are conserved quantities.}
\label{knot}
\end{center}
\end{figure}

We consider tubular knots as tubular embeddings of the magnetic field 
in an ideal, incompressible, perfectly conducting fluid in $S^{3}$
(i.e. $\RR^{3}\cup\{\infty\}$, simply connected). The magnetic field 
$\bB=\bB(\bx,t)$ ($\bx$ the position vector and $t$ time) 
is frozen in the fluid and has finite energy, that is
\begin{equation}
	\bB\in\{\nab\cdot\bB=0, \ 
	\partial_t\bB=\nab\times(\bu\times\bB), \ 
	L_{2}{\rm -norm} \}\ .
	\label{field}
\end{equation}
A magnetic knot $K$ is given by the embedding of the magnetic field in
a \emph{regular} tubular neighbourhood $\cT_{a}$ of radius $a>0$,
centred on the knot axis $\cC$ of local radius of curvature $\rho>0$ (see
Figure \ref{knot}).  The field is actually embedded onto nested tori
$\cT_{i}$ $(i=1,\ldots,n)$ in $\cT_{a}$, and regularity is ensured by
taking $a\le\rho$ pointwise along $\cC$.  The existence of
non-self-intersecting nested tori in $\cT_{a}$ is guaranteed by the
tubular neighbourhood theorem (Spivak, 1979).  $\cC$ is assumed to be
a $C^{2}$-smooth, closed loop (submanifold of $S^{3}$ homeomorphic to
$S^{1}$), simple (i.e. non-self-intersecting) and parametrized by
arc-length.  The total length of $\cC$ is $L=L(\cC)$.  Evidently $K$
has the knot type of $\cC$, being either \emph{trivial}, if $\cC$ (an
unknot) bounds a smoothly embedded disk, or \emph{essential}.  For
simplicity we take $\cT_{a}=\cC\otimes\cS$ given by the product of
$\cC$ with the solid circular disk $\cS$ of area $A=\pi a^{2}$, taken
in the cross-sectional plane to $\cC$.  The total volume
$V=V(\cT_{a})=\pi a^{2}L$.  We assume the tubular boundary
$\partial\cT_{a}=\partial\cT$ (dropping the suffix) remain a magnetic
circular, cylindrical surface at all times, of uniform cross-section
all along $\cC$; denoting by $\bnu_{\perp}$ the unit normal to
$\partial\cT$, we have $\textrm{supp}(\bB):=K$, where
\begin{equation}
        \cT(K)\;\hookrightarrow\;S^{3}\ , \qquad \textrm{with}\qquad 
	\bB\cdot\bnu_{\perp}=0\ 
	\textrm{on}\ \partial\cT\ .
	\label{magneticknot}
\end{equation}

The magnetic flux $\Phi$ through the cross-section $\cS$ is given by
\begin{equation}
	\Phi=\int_{A(\cS)}\bB\cdot\bnu\,\rd^{2}\bx\ ,
	\label{flux}
\end{equation}
where now $\bnu$ is the unit normal to $\cS$.
In ideal conditions the knot $K$ evolves under the action of the group 
of volume- and flux-preserving diffeomorphisms $\varphi:\; K\to 
K_{\varphi}$. Magnetic energy $M(t)$ and 
magnetic helicity $H(t)$ are two fundamental physical quantities, 
defined by:
\begin{equation}
	M(t)=\frac12\int_{V(K)}\!\! \|\bB\|^2 \, \rd^3\bx\ ,
	\label{e}
\end{equation}
and
\begin{equation}
	H(t)=\int_{V(K)}\!\! \bA\cdot\bB \, \rd^3\bx\ ,
	\label{h}
\end{equation}
where $\bA$ is the vector (Coulomb) potential associated with 
$\bB=\nab\times\bA$. As usual, we take $\nab\cdot\bA=0$ in $S^3$.  
For frozen fields helicity is a conserved quantity (Woltjer, 1958),
thus $H(t)=H=\,$constant. Moreover, it is well-known that helicity 
admits topological interpretation in terms of linking numbers, and 
for a single magnetic knot we have (Berger \& Field, 1984; Moffatt 
\& Ricca, 1992):

\vskip4pt plus2pt
\noindent
{\bf Theorem 1.} {\em Let $K$ be an essential magnetic knot in an
ideal fluid. Then} 
\begin{equation}
	H=Lk\,\Phi^2=(Wr+Tw)\Phi^2 \ ,
	\label{hel}
\end{equation}
{\em where $Lk$ denotes the C\u alug\u areanu-White linking
number of $\cC$ with respect to the framing induced by the 
embedding of the $\bB$-lines in $\cT$.}     
\vskip4pt plus2pt

The two \emph{geometric} quantities $Wr$ and $Tw$ are the writhing 
number and the twist number: $Wr$ is a measure of the average coiling 
and distortion of $\cC$ in space and depends only on the geometry of 
$\cC$, while $Tw$ measures the winding of the field-lines around $\cC$, 
thus depending on the embedding of the $\bB$-lines within $\cT$. 
\emph{Zero-framing} of the field lines denotes zero-linking ($Lk=0$) 
of these lines with $\cC$, providing a reference measure for helicity 
calculations (see Appendix).

\section{Curvilinear coordinate system}
\label{sec:CCS}
\begin{figure}[t]
\begin{center}  
\epsfysize=70mm\epsfbox{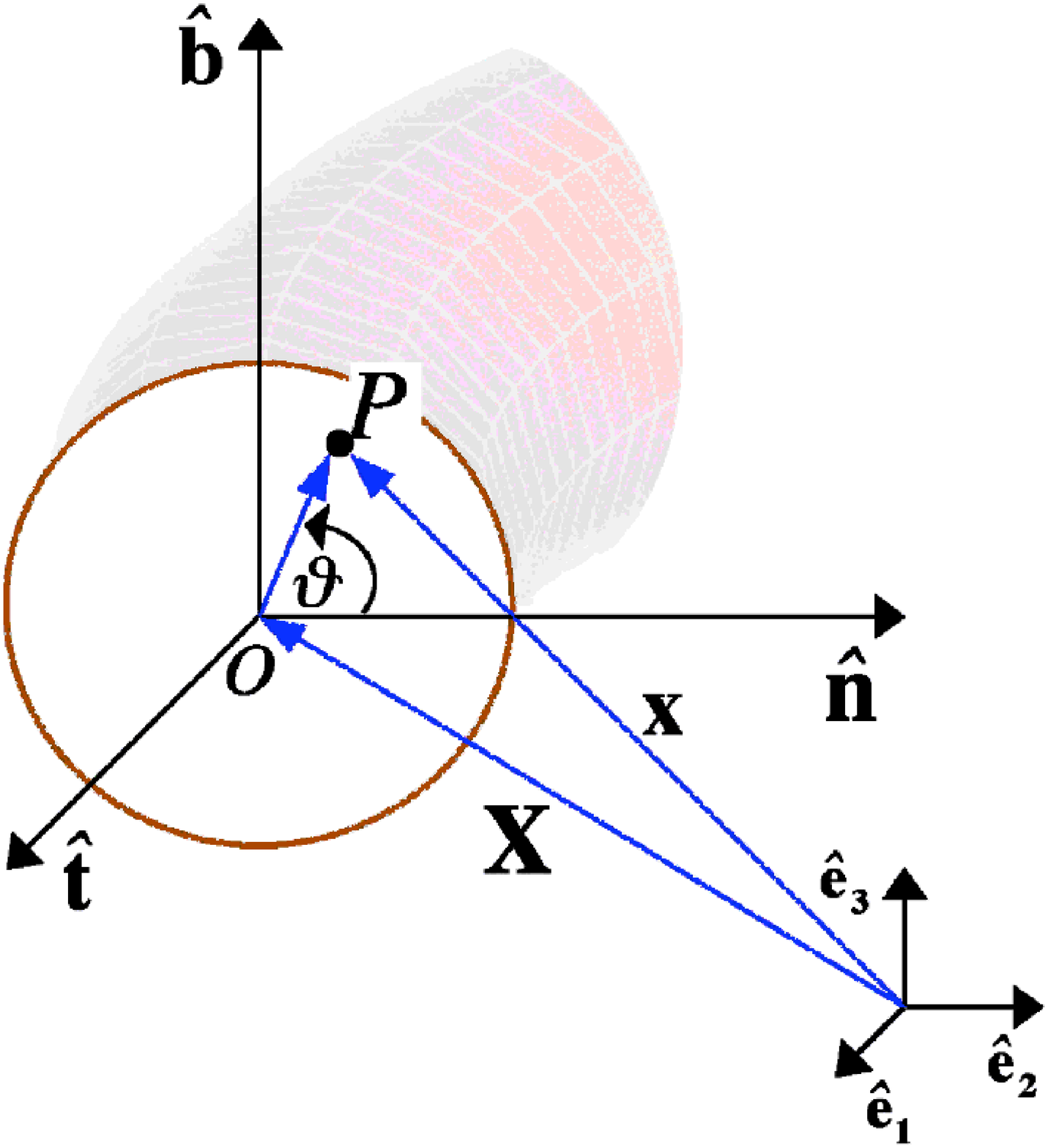}
\caption{Relationship between fixed reference frame and the Frenet 
frame $\{\bt,\bn,\bb\}$ for a point $P$ in the tube
cross-section $\cS$.}
\label{reference}
\end{center}
\end{figure}

It is useful to adopt an orthogonal curvilinear system of coordinates 
centred on $\cC$. Let $\cC$ be parametrized by the equation $\bx=\bX(s)$, 
where $s$ is arc-length, with origin $s=0$ at some point $O\in\cC$. 
Let $\bt(s)=\bX^{\prime}(s)$ be the unit tangent to $\cC$, prime denoting 
arc-length derivative. We take $\cC$ to be inflexion-free, then $\bn(s)$ and 
$\bb(s)$ are respectively the standard unit normal and binormal to $\cC$, 
with curvature $c=c(s)=1/\rho(s)$ and torsion $\tau=\tau(s)$ given by the standard 
Frenet-Serret equations, i.e.
\begin{equation}
    {\bt^{\prime}}=c\bn\ ,\quad 
    {\bn^{\prime}}=-c\bt+\tau\bb\ ,\quad
    {\bb^{\prime}}=-\tau\bn \ .
    \label{FrenetSerret}
\end{equation}
If $c(s)=0$ at some point of $\cC$, then $\cC$ has there an inflexion
and $\bn(s)$ is undefined.  Since $\cC$ is assumed to be
$C^{2}$-smooth, local inflexional configurations will be resolved by a
continuity argument from either sides of the inflexion point.  Thus,
without loss of generality, we shall take $\cC$ to be inflexion-free,
that is $c(s)>0$, $\forall s\in[0,L]$, so that the Frenet triad
$\{\bt,\bn,\bb\}$ is everywhere well-defined on $\cC$.
A point $P\in\cS$ (see Figure \ref{reference}) is thus given by
\begin{equation}
\bx=\bX(s)+r\cos\vartheta\bn(s)+r\sin\vartheta\bb(s)\ ,
\label{pointP}
\end{equation}
where $(r,\vartheta)$ is the polar coordinate system in the
cross-sectional plane. 

\vskip4pt plus2pt
\noindent
{\bf Lemma 2.} {\em Let $\cT$ be a regular tubular neighbourhood of 
$\cC$. Then the system of coordinates $(r,\thr, s)$, where
\begin{equation}
     \thr=\vartheta+\int_{0}^{s}\tau(\bar s)\,\rd\bar s\ ,
\label{thetaslemma}  
\end{equation}
provides a zero-twist reference system on $\cC$ in $\cT$.}     
\vskip4pt plus2pt

\noindent
\emph{Proof.} Let us consider a second curve $\cC^{\ast}$, 
of equation 
\begin{equation}
\bX^{\ast}(s)=\bX(s)+r\bTn(s) \ ,
\label{xstar}
\end{equation}
parallel transfer of $\cC$ at a distance $r$ in the normal direction 
along $\bTn=\bn\cos\vartheta^{\ast}+\bb\sin\vartheta^{\ast}$;
$\cC^{\ast}$ is a ``push-off'' of $\cC$ in the normal direction $\bTn$. 
The pair $\{\cC,\cC^{\ast}\}$ identifies the ribbon 
$\cR(\cC,\cC^{\ast})$ of edges $\bX(s)$ and $\bX^{\ast}(s)$, whose
twist provides a measure of the winding of $\cC^{\ast}$ around $\cC$.
For a generic infinitesimal displacement of the point $P\in\cS$ 
(see Figure \ref{reference}) in space, we have
\begin{equation}
\rd\bx=\left(\bt+\xi_{2}\frac{\rd\bTn}{\rd s}
        +\xi_{3}\frac{\rd\bTb}{\rd s}\right)\rd s
	+\bTn\rd\xi_{2}+\bTb\rd\xi_{3}\ ,
\label{iincrement}
\end{equation}
where $\xi_2$, $\xi_3$ denote coordinates along $\bTn$ and 
$\bTb=-\bn\sin\vartheta^{\ast}+\bb\cos\vartheta^{\ast}$,
respectively. Note that the triple $\{\bt,\bTn,\bTb\}$ is
orthogonal. By using (\ref{FrenetSerret}), we have 
\begin{equation}
\rd\bx =\left[(1-\xi_{2}c\cos\vartheta^{\ast}
          +\xi_{3}c\sin\vartheta^{\ast})\bt
	  +\left(\tau +\frac{\rd\vartheta^{\ast}}{\rd s}\right)
	  \left(\xi_{2}\bTb-\xi_{3}\bTn\right)\right]\rd s
	  +\bTn\rd\xi_{2}+\bTb\rd\xi_{3}\ .
\label{rdbx}
\end{equation}
Hence, the metric is given by
\begin{eqnarray}
\rd\bx\cdot\rd\bx&=&\left[\left(1-\xi_{2}c\cos\vartheta^{\ast}
                    +\xi_{3}c\sin\vartheta^{\ast}\right)^2
		    +\left(\xi_{2}^{2}+\xi_{3}^{2}\right)
    \left(\tau+\frac{\rd\vartheta^{\ast}}{\rd s}\right)\right]
		    \left(\rd s\right)^2 \nonumber\\
		 &&+\left(\rd \xi_2\right)^2
		    +\left(\rd \xi_3\right)^2
        +2\left(\tau+\frac{\rd \vartheta^{\ast}}{\rd s}\right)
		    \left(\xi_2\rd\xi_3-\xi_3\rd\xi_2\right)\ .
\label{gg}
\end{eqnarray}
The metric is orthogonal in the following cases:
\begin{equation}
    \textrm{(i)}\qquad\tau+\frac{\rd \vartheta^{\ast}}{\rd s}=0\ ;
    \qquad
    \textrm{(ii)}\qquad \xi_2\rd\xi_3-\xi_3\rd\xi_2=0\ .
    \label{ortho}
\end{equation}
Condition (ii) corresponds to a degenerate system of coordinates, 
since $\xi_2=\xi_3+\textrm{cst.}$;  condition (i) provides 
zero twist of the reference system $\{\bt,\bTn,\bTb\}$ everywhere 
along $\cC$, since (cf. Moffatt \& Ricca, 1992)
\begin{equation}
    Tw=\frac1{2\pi}\int_{\cC}\left(\tau(s)+\frac{\rd 
    \vartheta^{\ast}(s)}{\rd s}\right)\,\rd s \ .
    \label{tw}
\end{equation}
Integration of (\ref{ortho})--(i) gives
\begin{equation}
      \vartheta^{\ast}(s)=-\int_{0}^{s}\tau(\bar s)\rd\bar s+\vartheta_0\ ;
\label{thetast}	 
\end{equation}
without loss of generality we take $\vartheta_{0}=0$, so that
$\cR(\cC,\cC^{\ast})$ spans pointwise in the normal direction given 
by the push-off of $\cC^{\ast}$; the pair $\{\cC,\cC^{\ast}\}$ is 
untwisted and provides \emph{zero-twist} for the coordinate system. 
Hence, the metric is orthogonal and it is given by
\begin{equation}
    \rd\bx\cdot\rd\bx=[1-c(\xi_2\cos\vartheta^{\ast}-\xi_3 \sin
    \vartheta^{\ast})]^2(\rd s)^{2}+(\rd\xi_2)^2+(\rd\xi_3)^{2}\ ,
\label{metric}
\end{equation}
and by taking 
\begin{equation}
    \xi_2=r\cos\thr\ ,\qquad \xi_3=r\sin\thr\ ,
\label{polar}
\end{equation}
we have
\begin{equation}
       \rd\bx\cdot\rd\bx=[1-cr\cos(\thr+\vartheta^{\ast})]^2(\rd s)^2
       +(\rd r)^2+r^2(\rd\thr)^2\ .
\label{mercier}
\end{equation}
By using (\ref{thetast}), the independent coordinate $\vartheta_R$ is 
related to the polar angle $\vartheta$ by
\begin{equation}
     \thr=\vartheta+\int_{0}^{s}\tau(\bar s)\, \rd\bar s\ .
\label{thetas}  
\end{equation}
\par\hfill$\Box$

\noindent
The orthogonal system $(r,\thr, s)$ was originally found by Mercier (1963). 
In the orthonormal basis $\{\ber,\bethr,\bt\}$ the
scale factors $h_{1}$, $h_{2}$, $h_{3}$ are given by
\begin{equation}
    \left\{
    \begin{array}{l}
\displaystyle	
    \frac{\partial\bx}{\partial r}=h_{1}\ber
               =\cos\vartheta(\thr,s)\bn(s)+\sin\vartheta(\thr,s)\bb(s)
           \ ,\nonumber\\[3mm]
\displaystyle	   
    \frac{\partial\bx}{\partial\thr}=h_{2}\bethr
               =-r\sin\vartheta(\thr,s)\bn(s)+r\cos\vartheta(\thr,s)\bb(s)
           \ ,\nonumber\\[3mm]
\displaystyle	   
    \frac{\partial\bx}{\partial s}=h_{3}\bt
               =(1-c(s)r\cos\vartheta(\thr,s))\bt(s)\ .     
    \end{array}    
    \right.   
    \label{factors} 
\end{equation}
By setting $k=k(r,\thr,s)=1-c(s)r\cos\vartheta(\thr,s)$, the metric tensor 
is given by
\begin{displaymath}
    (g_{i,j})=(h_{i}\hat{\mathbf{e}}_{i})\cdot 
    (h_{j}\hat{\mathbf{e}}_{j}) =\left(
    \begin{array}{ccc}
           1 & 0 & 0\\
           0 & r^{2} & 0 \\
           0 & 0 & k^2 
    \end{array}
    \right)\ ,
\end{displaymath}
where $\hat{\mathbf{e}}_{1}=\ber$, $\hat{\mathbf{e}}_{2}=\bethr$, 
$\hat{\mathbf{e}}_{3}=\bt$, with determinant $g=r^{2}k^2$, and Jacobian 
\begin{equation}
J=(h_1\ber\cdot h_2\bethr)\times h_3\bt=\sqrt{g}= rk\ .
    \label{Jacobian}
\end{equation}
Since we assume that $r<c^{-1}=\rho$, we have $J>0$ so that the
transformation results well-defined. Hence, eq. (\ref{pointP}) 
reduces to 
\begin{equation}
     \bx=\bX(s) + r\ber\ , \quad\textrm{with}\quad
     \vartheta=\thr-\int_{0}^{s}\tau(\bar s)\, \rd\bar s\ .
\label{point}
\end{equation}

\section{Magnetic field and flux prescription}
Since $\partial\cT$ is a magnetic surface, the magnetic field $\bB$ is 
given purely in terms of poloidal and toroidal components, that is 
\begin{equation}
\bB=\bB_P + \bB_T=B_\thr\bethr+B_{s}\bt\ .
\label{Bfield}
\end{equation}
Moreover, since $\bB$ is divergenceless, we have
\begin{equation}
\nabla\cdot\bB=
  \frac{1}{rk}\left[\frac{\partial( k B_\thr)}{\partial\thr} 
   +\frac{\partial(r B_s)}{\partial s}\right] =0\ ,
\label{divergenceless}
\end{equation}
so that there exists a flux function $\psi(r,\thr,s)$, such that
\begin{equation}
B_\thr=\frac{1}{k}\frac{\partial\psi}{\partial s}\ ,\qquad
B_s=-\frac{1}{r} \frac{\partial\psi}{\partial\thr}\ .
\label{Bflux}
\end{equation}
For the magnetic field to be single-valued, $\psi(r,\thr,s)$ 
must be the sum of terms that are either linear or periodic in 
$\thr$ and $s$ (cf. Bateman 1978, pp. 127-128); hence, periodicity 
in both coordinates may be prescribed.

\begin{figure}[t]
\begin{center}  
\epsfysize=50mm\epsfbox{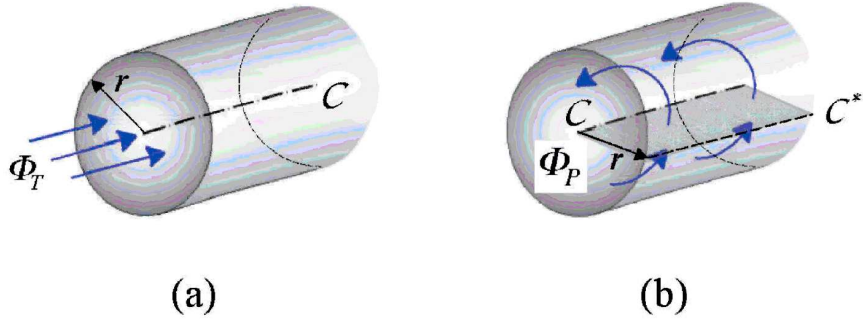}
\caption{(a) Toroidal and (b) poloidal flux in $\cT$.}
\label{fluxes}
\end{center}
\end{figure}

We now need to express the magnetic field in terms of toroidal and
poloidal flux.  Let $\Phi_{T}$ be the toroidal flux through a
cross-sectional area of $\cT$ of radius $r$ (see Figure
\ref{fluxes}-(a)):
\begin{equation}
\Phi_{T}=\int_{0}^{r}\int_0^{2\pi} B_s \bar r\, {\rd}\thr\, {\rd} \bar r\ .
\label{toroidal}
\end{equation}
From the second of (\ref{Bflux}) and the periodicity of $\psi$ in 
$\thr$, we have
\begin{equation}
    \Phi_{T}(r)=-2 \pi\int_0^{r}\frac{\partial\psi}{\partial\thr}\,{\rd}\bar r \ ,
    \label{psit}
\end{equation}
hence
\begin{equation}
\psi(r,\thr,s) =-\frac{\thr}{2\pi}\frac{\rd\Phi_{T}(r)}{\rd r}+c_1(r,s) \ .
\label{psione}
\end{equation}
The poloidal flux $\Phi_{P}$ through a ribbon area of width $r$
(see Figure \ref{fluxes}-(b))  is given by 
\begin{equation}
\Phi_{P}=\int_{0}^{L/k}\!\!\int_{0}^{r} B_{\thr}\, {\rd}\bar r\, k{\rd} s  \ .
\label{poloidal}
\end{equation}
From the first of (\ref{Bflux}) and the periodicity of $\psi$ in $s$, we have
\begin{equation}
    \Phi_{P}(r)=\int_{0}^{r}\frac{L}{k}\frac{\partial\psi}{\partial 
    s}\,{\rd}\bar r \ ,
    \label{psip}
\end{equation}
so that
\begin{equation}
\psi(r,\thr,s)=\frac{s}L\frac{\rd \Phi_{P}(r)}{\rd r}+c_2(r,\thr)\ .
\label{psitwo}
\end{equation}
Thus, from (\ref{psione}) and (\ref{psitwo}), we have
\begin{equation}
\psi(r,\thr,s)=-\frac{\thr}{2\pi}\frac{\rd\Phi_{T}(r)}{\rd r} 
               +\frac{s}{L}\frac{\rd\Phi_{P}(r)}{\rd r}
               +\widetilde{\psi}(r,\thr,s)\ ,
\label{psitot}	       
\end{equation}
where $\widetilde{\psi}=\widetilde{\psi}(r,\thr,s)=c_1(r,s)+c_2(r,\thr)$ 
is a single-valued function, periodic in $\thr$ (with period $2\pi$) 
and in $s$ (with period $L$). By eqs. (\ref{Bflux}), we have 
\begin{equation}
\bB=\left(0,\frac{1}{L}\frac{\rd\Phi_{P}}{\rd r},
     \frac{1}{2\pi r}\frac{\rd\Phi_{T}}{\rd r}\right)
     +\left(0,\frac{\partial\widetilde{\psi}}{\partial s},
     -\frac{\partial\widetilde{\psi}}{\partial \thr}\right)\ ;
\label{B}
\end{equation}
the total field is given by the sum of an average field, represented 
by the first term, plus a fluctuating field with zero net flux.  

\section{Constrained minimization of magnetic energy}
\label{constrained}
\subsection{Standard flux tube}
Let us specify the relation between toroidal and poloidal flux.
Let $V_r=\pi r^{2}L$ be the partial volume of the tubular neighbourhood of 
radius $r$; the ratio of the partial to total volume is given by 
$V_{r}/V(\cT)=(r/a)^{2}$. Now, let $f(r/a)$ be a 
monotonically increasing function of $r/a$; for example 
\begin{equation}
    f(r/a)=\left(\frac{r}{a}\right)^{\gamma}\ ,\qquad 
    (\gamma>0)\ ;
\label{qfactor}	
\end{equation}
$\gamma=2$ gives the standard ratio of partial to total volume.
Denoting by $\Phi:=\Phi_T\left(a\right)$ the total flux, we have
\begin{equation}
\Phi_{T}(r)=\left(\frac{r}{a}\right)^{\gamma}\Phi\ ,\qquad
\Phi_{P}(r)=h\left(\frac{r}{a}\right)^{\gamma}\Phi \ ,
\label{phitp}
\end{equation}
where $h$ denotes the \emph{magnetic field framing}, given by
$(2\pi)^{-1}$ times the turns of twist required to generate the
poloidal field from the toroidal field, starting from $\Phi_P=0$.
A direct calculation of helicity in terms of fluxes shows that 
$h$ is indeed the linking number $Lk$ of the embedded 
field (see Appendix). A \emph{standard} flux tube (cf. CM95) is 
defined by taking $\gamma=2$.

\subsection{Topological bounds on energy minima}
Relaxation of magnetic knots under topological constraints 
has been studied by several
authors, including Arnold (1974), Moffatt (1990), Freedman (1988),
Freedman \& He (1991).  Various bounds on magnetic energy $M(t)$ and
relationships between energy minima $M_{\rm min}$ and topological
complexity of knot type were found by these authors.  In particular,
for zero-framed knots, Ricca (2008) has proven that:

\vskip4pt plus2pt
\noindent
{\bf Theorem 3.} {\em Let $K$ be a zero-framed, essential magnetic 
knot, embedded in an ideal fluid. Then, we have}
\begin{equation}
  {\rm (i)}\quad 
     M(t)\ge \left(\frac{2}{\pi}\right)^{1/3}
     \frac{\vert H\vert}{{V}^{1/3}}=0 \ ;\qquad 
  {\rm (ii)}\quad 
     M_{\rm min}=\left(\frac{2}{\pi}\right)^{1/3}
     \frac{\Phi^{2}}{V^{1/3}}c_{\rm min}\ ,
\label{MM}	
\end{equation}
{\em where $c_{\rm min}$ is the topological crossing number of $K$.}
\vskip4pt plus2pt

\noindent
Note: results of Theorem 3 refer to the definition of magnetic 
energy given by eq. (\ref{e}), that involves the coefficient $1/2$.

\begin{figure}[t]
\begin{center}  
\epsfysize=60mm\epsfbox{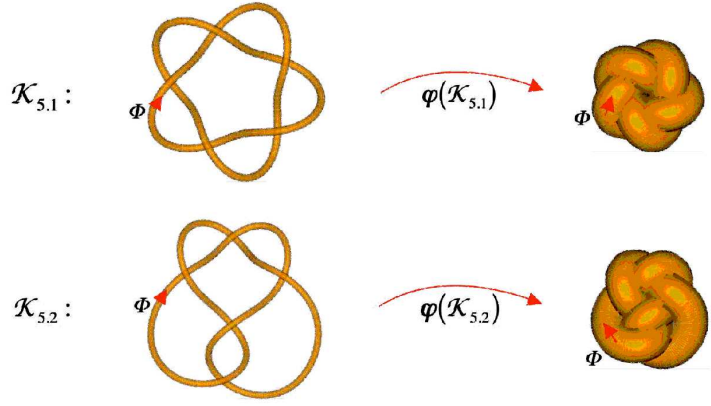}
\caption{On the basis of eq. (\ref{MM})-(ii) two different 
zero-framed knot types with same volume, flux and $c_{\rm min}$
(for example the $K_{5.1}$ and $K_{5.2}$ shown, with $c_{\rm min}=5$), 
may both relax to the same minimum energy.}
\label{relaxation}
\end{center}
\end{figure}

\par
By Theorem 1 $Lk=0$ yields $H=0$, hence for zero-framed knots the
lower bound given by inequality (\ref{MM})-(i) simply reduces to state
the positiveness of $M(t)$.  Equation (\ref{MM})-(ii), however, puts
in relation knot topology and magnetic energy minima through
crossing number information.  By comparing eq.  (\ref{MM})-(ii) with
(\ref{moffatt}) we see that $c_{\rm min}=m(0)$.  Relationship
(\ref{MM})-(ii), though, is not exhaustive, since with the exception of
the $3$- and the $4$-crossing knot, all other isotopes have several
topologically distinct representatives (see section \S \ref{last}
below).  Hence, inequality (\ref{MM})-(ii) cannot help to detect knot
types of same $c_{\rm min}$ on the mere basis of their minimum energy
level (see the example of Figure \ref{relaxation}).  For these cases
specific other detectors are necessary.

\subsection{Constrained relaxation to groundstate energy}
\label{constrelaxation}
We now look for groundstate energy of magnetic knots by the minimization of 
magnetic energy (\ref{e}) under specific constraints, following the same 
approach of CM95. The minimum energy thus obtained provides an upper bound 
to the true minimum attainable in absence of constraints. Let $M^{\ast}$ 
and $L^{\ast}$ denote the constrained minimum magnetic energy and the 
corresponding minimal knot length. We have.

\vskip4pt plus2pt
\noindent
{\bf Theorem 4.} {\em Let $K$ be an essential magnetic knot in an ideal fluid,
with signature $\{V,\Phi\}$ and magnetic field given by (\ref{B}).
We assume that\newline
\indent (i) \ $\{V,\Phi\}$ is invariant;\newline
\indent  (ii) \ the circular cross-section $\cS$ is independent of the 
arc-length $s$;\newline
\indent (iii) \ $\widetilde{\psi}$ is independent of the arc-length 
$s$;\newline
\indent (iv) \ the knot length $L$ (or $L^{\ast}$) is independent of 
the knot framing $h$.
\newline
Minimization of magnetic energy, constrained by (i)-(iv), yields}
\begin{equation}
   M^{\ast}=\left(\frac{\gamma^2 {L^{\ast}}^2}{8(\gamma-1)V}
	      +\frac{\gamma\pi h^{2}}{2{L^{\ast}}}\right)\Phi^2\ .
\label{Mresult}	      
\end{equation}
\vskip4pt plus2pt

\noindent
\emph{Proof.} Let us express the magnetic energy (\ref{e}) in the 
orthogonal coordinates $(r,\thr, s)$:
\begin{equation}
M(t)=\frac12\int_0^{L/k}\!\!\int_0^{2\pi}\!\int_0^a
(B_\thr^2+B_{s}^2)\, {\rd}r\, r{\rd}\thr\, k{\rd} s\ .
\label{energyort}
\end{equation}
Since $\widetilde{\psi}$ is periodic in $\thr$, we define 
$\Lambda(r,\thr)=-\partial\widetilde{\psi}/\partial\thr$ so to
have:
\begin{equation}
\int_{0}^{2 \pi}\Lambda(r,\thr)\,\rd\thr=0 \ .
\label{Lambda}
\end{equation}
By assumption (iii) and eq. (\ref{Lambda}), we have:
\begin{eqnarray}
M(t)&=&\frac12\int_{0}^{L/k}\!\!\!\int_{0}^{2\pi}\!\!\int_0^a\!
    \left[\left(\frac{1}{L}\frac{{\rd}\Phi_P}{{\rd} r}\right)^2
     +\left(\frac{1}{ 2\pi r}\frac{\rd\Phi_T}{\rd 
     r}+\Lambda\right)^2\right] \,\rd r\, r\rd\thr\, k\rd s
     \nonumber \\
    &=&
    \frac12\int_{0}^{L/k}\!\!\!\int_{0}^{2\pi}\!\!\int_0^a\!
    (\Lambda^2+2P\Lambda+Q)\,\rd r\, r\rd\thr\, k\rd s\ ,
\label{men}
\end{eqnarray}
where
\begin{equation}
P=\frac{1}{2\pi r}\frac{\rd\Phi_T}{\rd r}\ ,\qquad
Q=\frac{1}{L^2}\left(\frac{\rd\Phi_P}{\rd r}\right)^2 
   +\frac{1}{4\pi^{2}r^2}\left(\frac{\rd\Phi_T}{\rd r}\right)^2\ .  
\label{lmn}
\end{equation}
Now we partially minimize $M$ with respect to $\Lambda$,
subject to the above constraints; let us introduce a Lagrange 
multiplier $\lambda(r)$ and minimize 
\begin{equation}
    \frac12(\Lambda^2+2P\Lambda+Q)-\lambda(r)\Lambda(r,\thr)
    \label{lagrange}
\end{equation}
with respect to $\Lambda$; we obtain
\begin{equation}
\Lambda=\lambda(r)-P\ ,\qquad
\int_{0}^{2\pi}\!(\lambda(r)-P)\,\rd\thr=0\ . 
\label{minlam}
\end{equation}
From the second of (\ref{minlam}), we have
\begin{equation}
    \lambda(r)=\frac1{2\pi}\int_0^{2\pi}P\,\rd\thr \ .
\label{lambda}
\end{equation}
Now, by substituting the first of (\ref{minlam}) into ($\ref{men}$), we 
get
\begin{equation}
\displaystyle{M^{\ast}=\frac12\int_{0}^{{L^{\ast}}/k}\!\!\!\int_{0}^{2\pi}
   \!\!\int_0^a\!
   \left(\lambda^2(r)-P^2+Q\right)\,\rd r\, r\rd\thr\, k\rd s\ ,}
\label{min1}   
\end{equation}
where $(\cdot)^\ast$ represents constrained minimization; from (\ref{lmn}) 
and ($\ref{lambda}$), we have
\begin{equation}
M^{\ast}=\frac12\int_{0}^{a}\left[\frac{{L^{\ast}}}{2\pi r}
         \left(\frac{\rd\Phi_T}{\rd r}\right)^2
	 +\frac{2\pi r}{{L^{\ast}}}
	 \left(\frac{\rd\Phi_P}{\rd r}\right)^2\right]\,\rd r \ .
\label{min2}   
\end{equation}
By taking $\Phi_P=h\Phi_T$ (cf. eq. \ref{phitp}), the integral (\ref{min2}) 
reduces to
\begin{equation}
   M^{\ast}=\int_{0}^{a}\left(\frac{\rd\Phi_T(r)}{\rd r}\right)^2 
   \frac{{L^{\ast}}^2 +4{\pi}^2 r^2 h^2}{4\pi r{L^{\ast}}}\,\rd r \ .
\label{min3} 
\end{equation}
Finally, by first of (\ref{phitp}) and straightforward integration, we have  
\begin{equation}
   M^{\ast}=\left(\frac{\gamma^2 {L^{\ast}}}{8\pi(\gamma-1)a^2}
              +\frac{\pi\gamma h^2}{2{L^{\ast}}}\right)\Phi^2 
	   =\left(\frac{\gamma^2 {L^{\ast}}^2}{8(\gamma-1)V}
	      +\frac{\gamma\pi h^{2}}{2{L^{\ast}}}\right)\Phi^2\ ,
\label{Mend}	      
\end{equation}
where, in the last term, we have substituted $a=(V/\pi {L^{\ast}})^{1/2}$ 
to express $M^{\ast}$ in terms of the signature $\{V,\Phi\}$. 
\par\hfill$\Box$

\noindent
As we see, the constrained minimum energy depends on the length axis
${L^{\ast}}$ and the square of the twist parameter $h$.  
For $\gamma=2$, eq. (\ref{Mend}) reduces further to
\begin{equation}
M^{\ast}=\left(\frac{{L^{\ast}}^2}{2V}
+\frac{\pi h^2}{{L^{\ast}}}\right)\Phi^2 \ .
\label{CM}
\end{equation}
At minimum energy, ${L^{\ast}}$ is the minimal tube length of the relaxed 
``tight'' knot.

\section{Constrained minimum energy by the SONO algorithm}
As we see from eq. (\ref{CM}), the constrained minimum energy of 
tubular knots is dictated by the tube minimal length. A useful 
measure of knot complexity is the knot \emph{ropelength} $L/R$, 
where from now on (dropping the suffix $(\cdot)^{\ast}$) $L$ and 
$R$ denote the minimal knot length and the radius of the maximal, 
circular cross-section of the tightest knot configuration. For 
simplicity let us set $V=1$ and $\Phi=1$. Under signature-preserving 
flow from $V=1=\pi R^{2}L$, by elementary algebra, we have
\begin{equation}
    1=\pi R^{3}\left(\frac{L}{R}\right)
    \qquad \Rightarrow \qquad 
    R=\left(\frac{1}{\pi(L/R)}\right)^{1/3}\ ,
    \label{R}
\end{equation}
so that we can express the tight knot length $L$ in terms of 
the ropelength $L/R$:
\begin{equation}
    L=(L/R)\left(\frac{1}{\pi(L/R)}\right)^{1/3}
            =\left(\frac{(L/R)^{2}}{\pi}\right)^{1/3}\ .
    \label{L}
\end{equation}
Equation (\ref{CM}) can thus be re-written purely in terms of 
ropelength and internal twist, that is 
\begin{equation}
M^{\ast}=\frac{(L/R)^{4/3}}{2\pi^{2/3}}
+\frac{\pi^{4/3}h^2}{(L/R)^{2/3}} \ .
\label{mr}
\end{equation}
We can now calculate $M^{\ast}$ by using algorithms that minimize
ropelength for any given internal twist value.  For this we shall rely
on one of the most advanced algorithms available, namely the SONO
algorithm, originally developed by Pieranski (1998) and collaborators.

\subsection{The SONO algorithm}
\label{SONO}
The Shrink--On--No--Overlaps (SONO) algorithm is a numerical software
implemented by Sylwester Przbyl (2001) under the direction of Piotr
Pieranski (Pieranski, 1998), and subsequently improved by other
collaborators.  Technical details of this algorithm are widely
available in literature; the interested reader can consult several
papers such as Pieranski \emph{et al.} (2001), Pieranski \& Przybyl
(2002), Baranska \emph{et al.} (2004), Baranska \emph{et al.} (2005).
We present here the ideas behind the algorithm and we highlight some
critical issues.

\subsubsection{The perfect rope model}
The tightening process is based on the idea that the tubular knot is
modelled by a perfectly flexible and infinitely hard, cylindrical
rope.  This means that the rope can be bent with zero force (no
elastic energy stored) and it cannot be squeezed.  This means that
the rope cross-section remains always perfectly circular.  It is also
assumed that the rope surface is perfectly frictionless and that the
knot axis, together with its tangent, is everywhere smooth and
continuous.  This guarantees that the perpendicular cross-section,
given by a disk centered on the rope axis, is everywhere always
well-defined.  Moreover, there is a control that enforces that
perpendicular cross-sections do not overlap to avoid
self-intersections.

\subsubsection{Discretization}
The knot axis is standardly discretized piecewise linearly by a finite
sequence of segments (beads) and vertices, to form a self-avoiding,
polygonal knot $K_{P}$ in $\RR^3$.  The tightening process starts from
a non-equilateral polygon and proceeds to an equilateral
configuration.  The main difficulty here is to produce a polygonal
knot, that best reproduces the knot axis of the perfect rope.  In
finding the tightest (or \emph{ideal}) configuration the initial
ropelength provides an upper bound for the minimal value of the ideal
configuration.  To simulate the hard shell of the perfect rope, hard
spheres are centred on the $K_P$ vertices.  Hence, the collection
(union) of all these spheres generates a corrugated rope surface.
This corrugation is made vanishingly small by increasing the number of
vertices.

\subsubsection{Original procedure and subsequent improvements}
SONO's basic goal is to minimize the length of $K_{P}$, while ensuring 
that (a) the rope cells remain well-defined at all times, and that 
(b) the rope cells do not overlap, while reducing $K_{P}$ to the equilateral 
polygon (a rope cell is made by a sphere with two opposite cups 
appropriately removed). To avoid overlapping of contiguous cells
a ``Control Curvature'' (CC) procedure is implemented to control 
(bound) the angle between consecutive beads. This is complemented by 
a second procedure (``Remove Overlaps'', RO), that controls the 
Euclidean distance between neighbouring (but not consecutive) vertices.
In order to produce an equilateral polygon a third procedure (``Equalize
Edges'', EE) is implemented. Hence, the tightening process is achieved by 
a an iterative application of the EE and RO procedures to minimize 
the differences in bead lengths and take account of emerging overlaps. 
The tightest configuration is thus given by an overlap-free, 
corrugated rope, centred on an equilateral polygonal knot.
A preliminary version of SONO (Przybyl, 2001) was followed by 
an improved version (Baranska \emph{et al.}, 2004), where a finer 
repositioning of the vertices is implemented through an appropriate 
displacement towards the local center of curvature. 

\subsection{The problem of local/global minimum}
The tightening process forces the rope to change its configuration. 
The end state is achieved when further shortening of the ropelength 
is no longer possible, because this would create non-removable overlaps. 
When the number of beads is small, the rope surface is highly 
corrugated, and, upon tightening, this could allow different part of 
the rope to be mutually entrenched in the grooves of the corrugation.
This would then give false information on the end state, and more 
specifically on local minima of the ropelength (read ``energy'').
As long as the number of cells is high, fine corrugation seems to 
prevent convergence to unrealistic local minima. However, at this 
stage a further readjustment of the vertices is not allowed and no
procedure is implemented to explore possible neighbouring minima.
Since global minimum configuration is not known \emph{a priori} 
(and no analytical criterium is available), there is no certainty 
that the minimal ropelength configuration realized by the SONO 
tightening process corresponds actually to a global minimum.

\section{Constrained minimum energy of tight torus knots: 
comparative results}
\begin{figure}[t]
\begin{center}  
\epsfysize=60mm\epsfbox{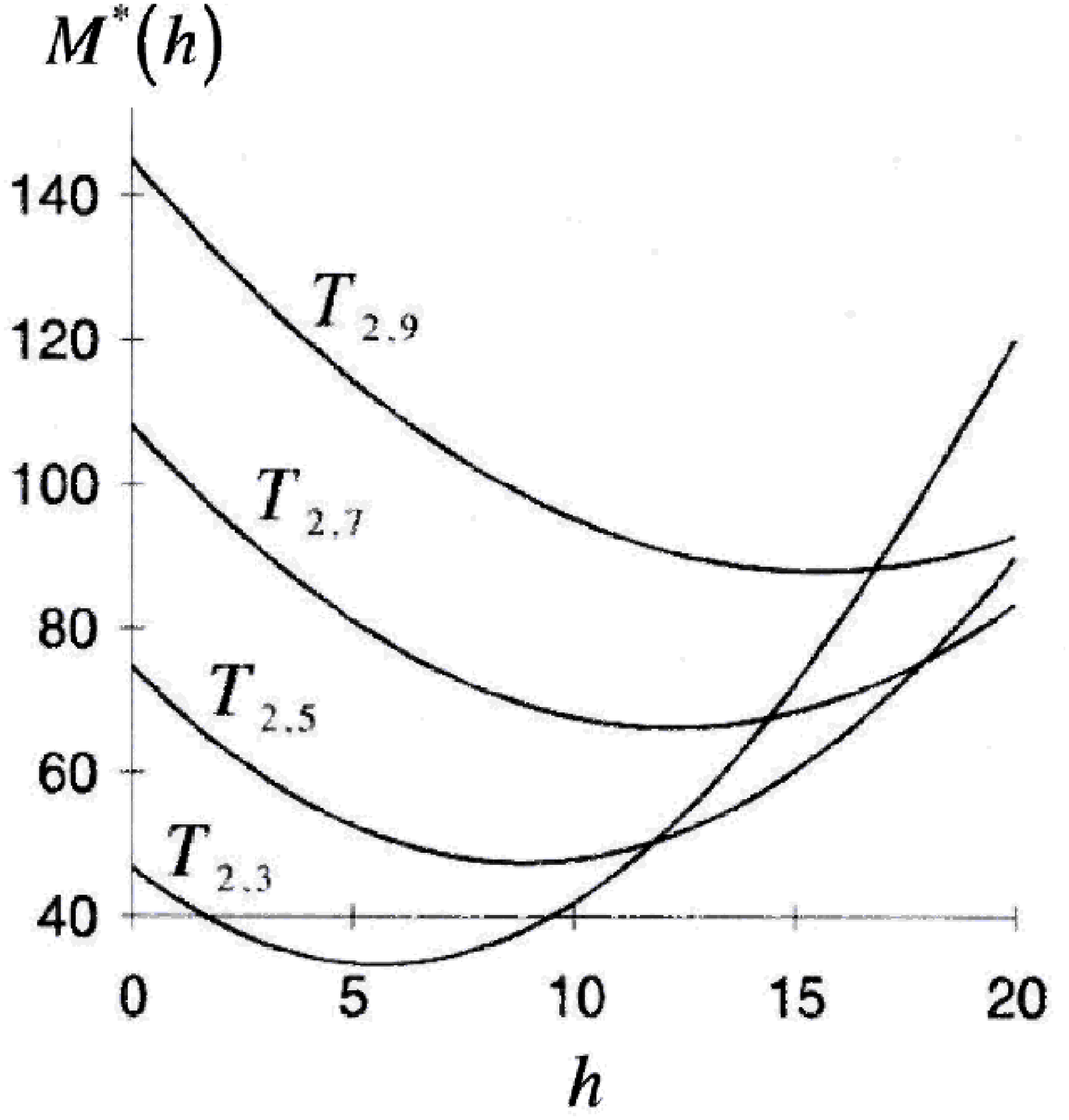}
\caption{Groundstate energy spectrum of torus knots obtained by Chui 
\& Moffatt (1995).}
\label{chui&moffatt}
\end{center}
\end{figure}

Orthogonal coordinates offer several advantages, one evidently being
the economy and transparency of the calculations involved.  Direct
comparison of eq.  (\ref{CM}) with eqs.  (7.1), (7.12) and (7.13) of
CM95 provides further evidence of this.  Before comparing results,
however, we should point out that eqs.  (5.19), (7.4), (7.9), (7.10)
and (7.13) of CM95 contains typos and errors that need correction: the
second integral in the r.h.s. of (7.10), for instance, must be
multiplied by $L$ to be correct (also on dimensional grounds!).  We
can then proceed with the comparison: let $\epsilon=a/L$; from eq.
(\ref{CM}), we have
\begin{equation}
   M^{\ast}=\frac1{\epsilon^{2}}\left(\frac{\gamma^2}{8\pi(\gamma-1)L}
              +\epsilon^{2}\frac{\pi\gamma h^2}{2L}\right)\Phi^2 \ ,
\label{Mend2}	      
\end{equation}
which, for $\gamma=2$, reduces to 
\begin{equation}
   M^{\ast}=\frac1{\epsilon^{2}}\left(\frac1{2\pi L}
              +\epsilon^{2}\frac{\pi h^2}{L}\right)\Phi^2 \ .
\label{CM2}	      
\end{equation}
By comparing eq. (\ref{CM2}) with eq. (7.1) of CM95 (supplied by (7.12) 
and the corrected (7.13)) we can see that the exact expression given 
by eq. (\ref{CM2}) corresponds to the contribution of (7.12) and of 
one term only of the corrected (7.13), with higher-order terms in 
the series expansion (7.1) left out. Here we should point out 
that the overall contribution of the cut-off terms, being these positive 
\emph{and} negative, may well sum up to zero. In any case, the result of 
(7.1) is then applied to study the groundstate 
energy of torus knot flux tubes (see \S 8 of CM95). Here the difficulty 
to analyse tight knot configurations was overcome by using standard 
torus knot equations (eq. (8.1) of CM95) and estimates on contact 
condition by some approximated form function (eq. (8.5) of CM95). The 
results obtained there are reproduced in Figure \ref{chui&moffatt}. 

In considering the relaxation to tight knot configuration, Chui \&
Moffatt made the implicit assumption that throughout the tightening
process the knot axis would have not changed its initial geometry, the
only change in time being an average increase of the tubular knot
circular cross-section.  However, only now we know that this
assumption is actually not legitimate.  Numerical implementation
of the global curvature concept to study tight knot configurations
(Gonzalez \& Maddocks, 1999) and extensive numerical simulations made
more recently by using the SONO algorithm (Pieranski \& Przybyl, 2002) 
show that the knot axis changes shape during the tightening process, 
with progressive deformation due to the change of curvature and torsion 
of $\cC$ from the initial configuration to the final end-state (while 
keeping the cross-section circular).

\begin{figure}[t]
\begin{center}  
\epsfysize=70mm\epsfbox{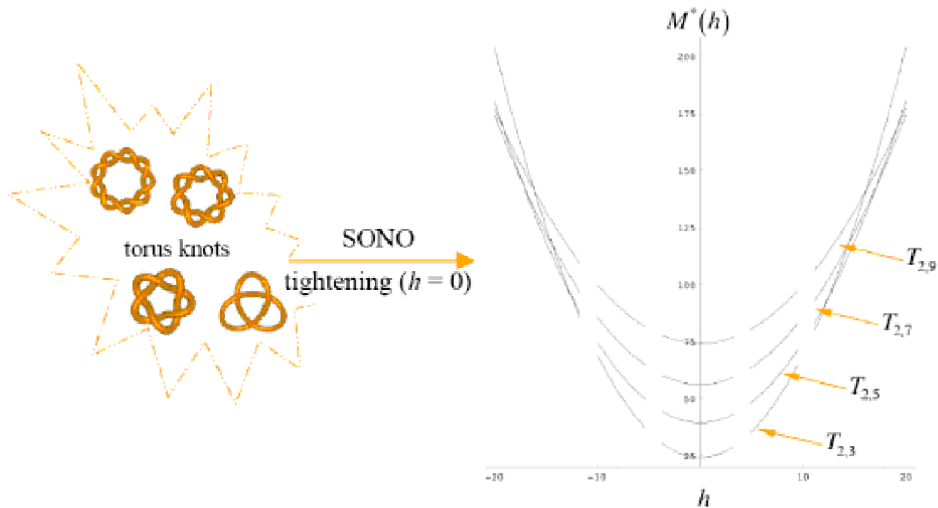}
\caption{Constrained energy minima of torus knots given by eq. 
(\ref{mr}). Data obtained by knot tightening performed by 
the SONO algorithm correspond to $h=0$ (Przybyl, 2001).}
\label{tighttorusknots}
\end{center}
\end{figure}

In the present case we can use the data on minimal ropelength of tight
torus knots $T_{p,q}$ ($p>q$; $p$, $q$ co-prime integers), obtained by
the SONO algorithm simulation (see Przybyl, 2001; pp.  47--49) for
$h=0$.  The results shown in Figure \ref{tighttorusknots} are given by
eq.  (\ref{mr}) obtained by setting $V=1$, $\Phi=1$ (consistently with
CM95).  The curves are based on the SONO data at $h=0$ and then
extrapolated according to eq.  (\ref{mr}).  Note that, by doing so, we
have assumed that minimal ropelength data remain independent from
$h$.  The quadratic dependence on the twist parameter $h$ is thus
evidenced by the family of parabolas.  For direct comparison with
results of CM95 (Figure \ref{chui&moffatt}), we report only the
constrained energy of torus knots of type $T_{2,q}$ ($q=3,\; 5,\; 7,\;
9$); similar parabolic curves (at higher energy levels) are obtained for
torus knots of higher topological complexity (see \S \ref{last}), such
as the $T_{3,q}$ ($q>3$), $T_{4,q}$ ($q>4$), and so forth. 

The two diagrams show some marked differences and some qualitative
similarities.  The first important difference is in the location of
the minima, that, in our case, are all centred at $h=0$.  Consistently
with eq.  (\ref{mr}), a non-zero framing (given by $h\neq 0$) yields
an increase of the energy level, due to the contribution of net twist
to the groundstate energy.  Moreover, take for example the energy
curve of the trefoil knot $T_{2,3}$ of Figure \ref{chui&moffatt}:
considering that the average writhing number of the tight
configuration is given by $Wr\approx 3.41$ (see, for example, the
writhing number values calculated by SONO), the value $h\approx 6$
(cf.  CM95, p.  626) attained at the minimum is hard to justify, even
by interpreting $h$ in terms of internal twist (since, by eqs.
(\ref{hel}), and \ref{helicitystandard} in Appendix, $h=0$ if and only
if $|Tw|=|Wr|$).  Moreover, the values at which the $T_{3,2}$,
topologically equivalent to the $T_{2,3}$, attains its minimum energy
(that with respect to $T_{2,3}$ are different in $h$ and $M^{\star}$)
seem to be completely wrong, considering that SONO tests on torus
knots demonstrated (Pieranski, 1998) that, for given $p$ and $q$,
$T_{q,p}$ and $T_{p,q}$ relax to the same tight configuration (hence,
same ropelength and energy).  Another difference is in the symmetry of
the diagrams.  The curves shown in Figure \ref{tighttorusknots} for
$h\neq 0$ are extrapolated from SONO data at $h=0$ according to eq.
(\ref{mr}).  The assumption we made on the independence of ropelength
from framing prevents us from stating that symmetry is actually
preserved when this assumption is relaxed.  Consistently with the
diagrams of CM95, though, the diagrams of Figure \ref{tighttorusknots}
show that above a certain internal twist threshold $h_{\rm cr}=h_{\rm
cr}(T_{p,q})$, minimum energy levels swap, reflecting the interplay
between ``internal'' and ``external'' topological complexity --- the
former being associated with the field twist and the latter with the
knot embedding in $S^{3}$.  At zero framing ($h=0$), however, internal
twist vanishes and energy levels relates to knot topology in good
agreement with (\ref{MM})--(ii).

\begin{figure}[t]
\begin{center}  
\epsfysize=70mm\epsfbox{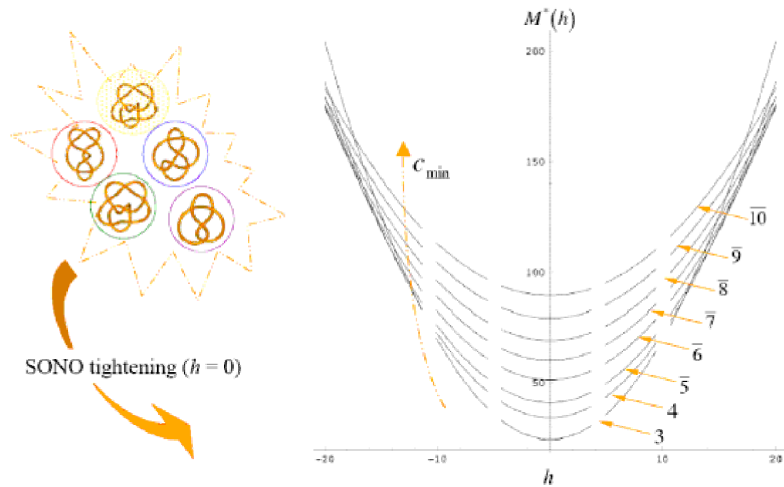}
\caption{Constrained energy minima of prime knots of
increasing complexity ($c_{\rm min}$) given by eq. (\ref{mr}). Data 
obtained by knot tightening performed by the SONO algorithm 
correspond to $h=0$ (Przybyl, 2001).  Overbar refers to data obtained 
by averaging values of tight knot lengths over the number of 
representatives within each family.}
\label{spectrum}
\end{center}
\end{figure}

\section{Constrained groundstate energy of prime knots up to 10 
crossings}
\label{last}
We can estimate the constrained energy minima of prime knots up to 10
crossings by using data obtained by the SONO tightening process (see
Przybyl, 2001; pp.  47--49).  In doing so, a word of caution is 
perhaps necessary: as remarked by Przybyl and Pieranski\footnote{See: 
\texttt{http://fizyka.phys.put.poznan.pl/\textasciitilde
pieransk/TablesUpTo9.html}}, one should bear in mind that the computed
tight knot values tabulated are not rigorous and are subject to 
computational improvements.
Assuming that the first decimal digit is correct, we inspect knot
types of increasing complexity given by their $c_{\rm min}$.  For 3
and 4 crossings there is only one knot type, but there are 2 for
$c_{\rm min}=5$, 3 for $c_{\rm min}=6$, 7 for $c_{\rm min}=7$, 21 for
$c_{\rm min}=8$, 49 for $c_{\rm min}=9$ and 166 for $c_{\rm min}=10$.
Since the distribution of the minimal length data produced by SONO is not,
apparently, a monotonic function of $c_{\rm min}$, the corresponding
energy levels result sometimes swapped between knot types of
increasing $c_{\rm min}$.  To analyse data effectively we proceed in
two steps.  First, from each family of given $c_{\rm min}$, we take
the average minimal length ($\bar{L}$) out of the total number of knot
representatives, and use this $\bar{L}$ to calculate the energy level of
that family.  The resulting curves, shown in Figure \ref{spectrum},
are therefore obtained by using eq. (\ref{mr}) and extrapolating the 
diagrams from the SONO minimal average ropelength data obtained at $h=0$.

\begin{figure}[t]
\begin{center}  
\epsfysize=80mm\epsfbox{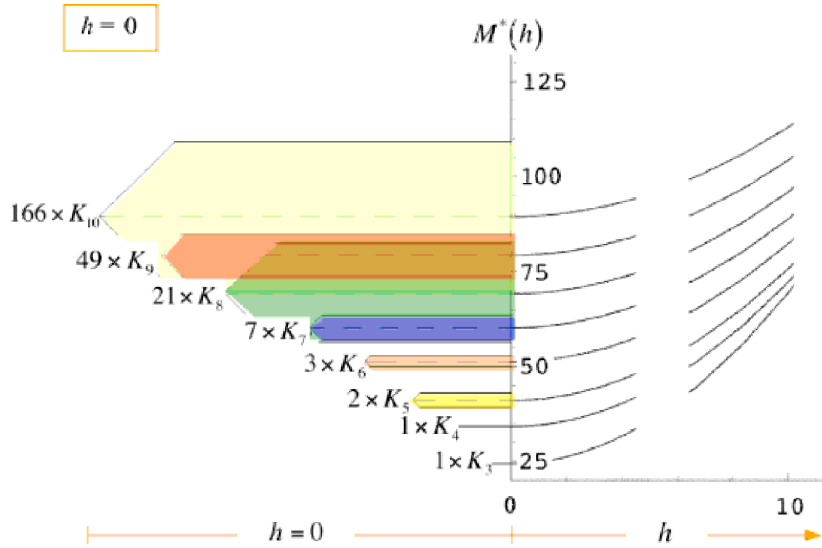}
\caption{Constrained energy minima of knot representatives 
(on the right) are compared with the corresponding bands of energy (on
the left) obtained for $h=0$.  Results are based on the data currently
available, obtained by the SONO tightening algorithm at $h=0$ (Przybyl, 2001).
Bands refer to the distribution of energy levels of knot
representatives belonging to a given knot family, denoted by
$K_{c_{\rm min}}$.  There is only one knot type for the 3 and the 4 crossing
families, whereas there are 3 for $c_{\rm min}=6$, 7 for $c_{\rm min}=7$, 21 
for $c_{\rm min}=8$, 49 for $c_{\rm min}=9$ and 166 for $c_{\rm min}=10$.}
\label{meanbands}
\end{center}
\end{figure}

To analyse the actual distribution of knots in relation to the
groundstate energy we present a second diagram (see Figure
\ref{meanbands}), where the median energy levels of Figure
\ref{spectrum} are reproduced on the right-hand-side for reference.
On the left-hand-side we report the actual distribution of energy
obtained for $h=0$, centred on the relative average value (dashed
lines).  Each band shows the spread of energy for each knot family of
given $c_{\rm min}$ (denoted by $K_{c_{\rm min}}$).  For $h\neq0$
these bands extend on either side of the vertical axis and are centred
on their parabolic median (not shown in Figure).  

\begin{figure}[th!]
\begin{center}  
\epsfysize=140mm\epsfbox{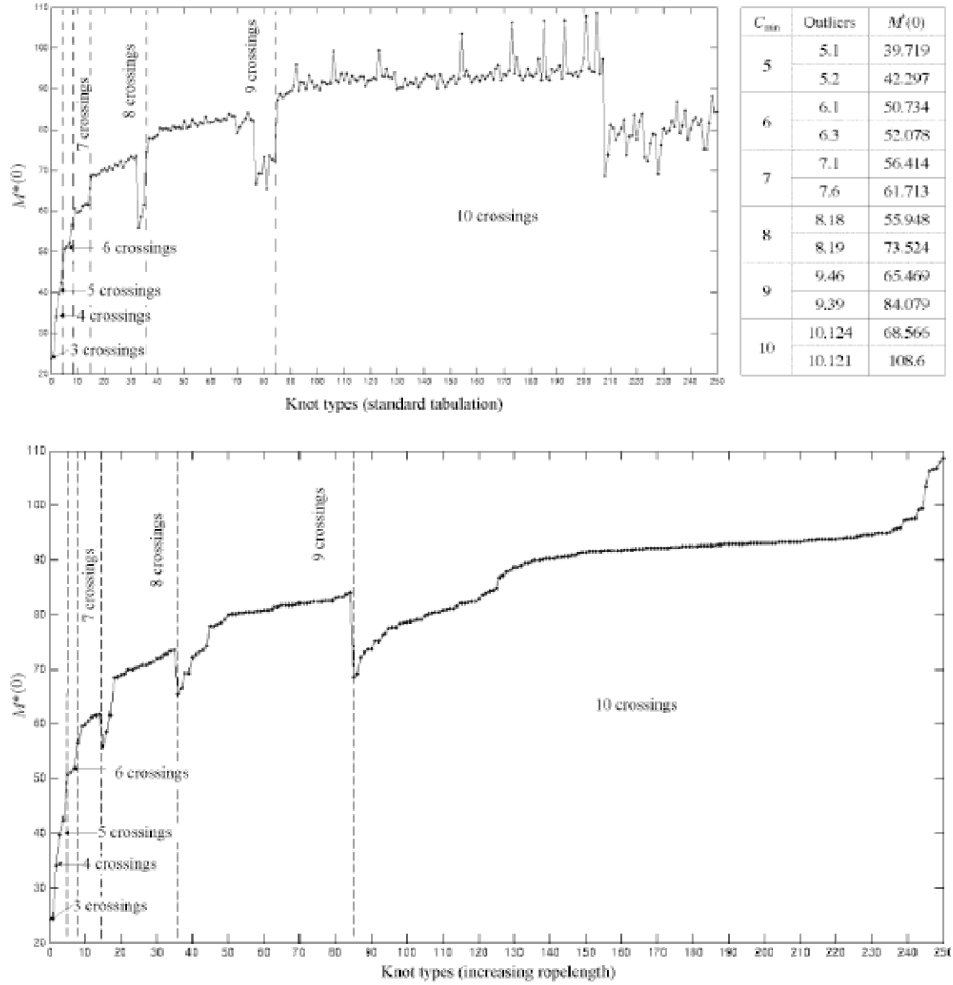}
\caption{Top diagram: constrained groundstate energy of tight knots up
to 10 crossings, based on data provided by the SONO algorithm; the
knots are distributed according to standard knot tabulation with
increasing crossing number.  The table on the right-hand-side lists
the knots (outliers) with minimum and maximum $M^{*}(0)$ ($M^{*}$ at
$h=0$) within each $c_{\rm min}$-family.  Bottom diagram: constrained
groundstate energy versus knot type; the knots are now distributed
according to their minimal ropelength values.}
\label{distribution}
\end{center}
\end{figure}

According to current SONO data, one-to-one
correspondence between constrained minimum energy and knot complexity
holds up to knot $K_{8.18}$ (i.e. it extends to include the first 32
knot types): the minimal length of $K_{8.19}$ appears to be shorter
than that of $K_{7.1}$.  It is curious to notice that, according to
standard (Rolfsen) knot tabulation, the knot representatives that 
break this correspondence are those (mostly non-alternating) listed 
in the last positions of the table: these are the last 3 in the 
8-crossing family, the last 8 in the 9-crossing family, and, with 
the exception of 3 knots, the last 40 in the $10$-crossing family. 
This is evidenced by the top diagram of Figure \ref{distribution},
that shows the energy distribution against knot types listed according
to the standard knot tabulation.  The table on the right-hand-side
lists the knot outliers (of each $c_{\rm min}$-family) with minimum
and maximum $M^{*}(h=0)$.  By re-ordering the knot types of each
$c_{\rm min}$-family in terms of increasing minimal ropelength we
obtain the bottom diagram of Figure \ref{distribution}, that clearly
demonstrates that, besides crossing number, ropelength too is a good 
detector of topological complexity.

\section{Concluding remarks}
\label{least}
The results presented in this paper on the groundstate energy of prime
knots up to 10 crossings are based on a new, exact analytical
expression for the constrained minima of magnetic energy.  This is
obtained as an improvement of previous work done by Chui \& Moffatt
(1995), by using appropriate orthogonal coordinates and standard
variational methods. A magnetic knot is identified with a tubular 
knot filled by a magnetic field, decomposed in poloidal and 
toroidal components. The tubular knot is given by a tube centred on 
the knot axis with circular cross-section, and the framing of the 
tubular knot is given by the linking of the field lines with the knot 
axis. By assuming volume and flux invariance and uniform, circular 
cross-section along the tube axis at all times, and independence 
of knot length from framing, we determine the constrained minimum 
magnetic energy functional (eq. \ref{Mresult}), and show that for 
standard flux-tube the minimizer reduces further to a simple expression 
(eq. \ref{mr}), function of ropelength and framing. Constrained 
minima are local minima for the magnetic relaxation process and 
provide an approximation from above to the actual minima. By using 
SONO data for tight knots we determine the groundstate energy 
spectrum of the first prime knots up to 10 crossings, and compare 
results for torus knots with previous work done by CM95. Comparison 
of diagrams show some marked differences and some qualitative 
similarities: contrary to CM95, in the present case the constrained 
groundstate energy reaches its minimum at zero-framing ($h=0$), and 
the minimizer results an even function of $h$. This means that the 
minimizer for a given framing and its negative are simply mirror 
images. 

Relaxation of basic constraints, numerical improvement of the
tightening process and improved numerical accuracy can all lead to
significant improvements in the energy bounds.  The assumption of
circular cross-section certainly represents the strongest limitation
to further minimization of energy and any relaxation of this
constraint is likely to bring appreciable improvements on the minima.
Evidently, this problem is paired with the difficulty of determining
the ``free volume'' of the interstices that survive to the
minimization, and since these are evidently related to the number of
crossings (Buck \& Simon, 1999), larger adjustments are expected for
knots of greater topological complexity.  Tight knot data may then be
sensitive to specific implementation of the tightening process.  Apart
from the test case of the trefoil knot (Gonzalez \& Maddocks, 1999),
there are no information on how different techniques, based on SONO,
global curvature radius, or else, compare.  Here too, significant
improvements of energy minima may be achieved when particularly
complex topologies are tested.  Finally, as work on trefoil knot
demonstrates (Baranska \emph{et al.}, 2004), computational progress in
numerical analysis (discretization techniques, error control, etc.)
improves accuracy and this, in turn, gets reflected in corrections to
energy levels.

From a more general viewpoint, then, a comparison between different 
techniques would be very beneficial. 
Existence of a global minimum in each isotopy class for minimizers
that depend solely on ropelength has been established by Cantarella
\emph{et al.} (2002), and an extension of this result to include
framing is, to the best of our knowledge, yet to be done. 
Alternative minimizers (for some ``twisto-elastic'' energy) that
depend on ropelength \emph{and} framing are not available, so that a
direct comparison with results obtained by different methods
is not possible.  The role of knot chirality, for example, that in the
present context is simply demanded to ropelength minimization, might
be detected by minimizers of different type.  This calls also for more
work by numerical simulations of knot tightening, for instance by
implementing new procedures not only for the fine tuning of local
minima, but also for investigating the role of framing, and for
inspecting symmetry issues in relaxed states.

In this sense the present work should be
considered as an important complement to current work on minimum
ropelength of tight knots (Litherland \emph{et al.}, 1999; Cantarella
\emph{et al.}, 2002) and properties of ideal knots (Gonzalez \& De La
Llave, 2003) and for applications, it has obvious implications for
all those problems in magnetohydrodynamics that involve estimates on
energy-complexity relations, especially in astrophysics, solar physics
and plasma physics.

\begin{flushleft}
\large{\bf Acknowledgements}
\end{flushleft}
We would like to thank Rob Kusner and P. Pieranski for providing some 
useful information and for updating us on current work on the SONO 
algorithm.

\begin{flushleft}
\Large{\bf Appendix}
\end{flushleft}
Magnetic helicity, given by (\ref{h}), can be written in terms of 
toroidal and poloidal flux. By using the orthogonal coordinates
$(r,\thr,s)$ we show that eq. (\ref{h}) reduces to eq. (\ref{hel})
with $Lk=h$.
In order to have $\Phi=0$ in $S^{3}\backslash\cT$ and helicity gauge-invariant, 
we take $\bA=(0,A_\thr,A_s)$ such that
\begin{equation}
\int_{0}^{L/k}\!\! A_s(a,0,s)\, k{\rd} s=0\ .
\label{condA}
\end{equation}
Since $\bB=(0,B_\thr,B_s)$, we must have
\begin{equation}
\bB=\nabla\times\bA=\frac{1}{r k}\det
\left(
\begin{array}{ccc}
\ber & r\bethr & k\bt \\[1mm]
\displaystyle{\frac{\partial}{\partial r}} &
\displaystyle{\frac{\partial}{\partial\thr}} &
\displaystyle{\frac{\partial}{\partial s}} \\[2mm]
0 & rA_\thr & k A_s 
\end{array}
\right)\ .
\label{matrix}
\end{equation}
From (\ref{B}), by equating corresponding components, we have 
\begin{eqnarray}
 \left\{ 
 \begin{array}{ll} 
  \displaystyle{
   \frac{1}{rk}\left[\frac{\partial(kA_s)}{\partial\thr}
   -\frac{\partial(rA_\thr)}{\partial s}\right]=0 }\ ,\\ [3mm]
  \displaystyle{-\frac{1}{k}\frac{\partial(kA_s)}{\partial r}
   =\frac{1}{L}\frac{\rd\Phi_P}{\rd r}
   +\frac{\partial\widetilde{\psi}}{\partial s} }\ ,\\ [3mm]
  \displaystyle{\frac{1}{r}\frac{\partial(rA_\thr)}{\partial r}
   =\frac{1}{ 2\pi r}\frac{\rd\Phi_T}{\rd r}
   -\frac{\partial\widetilde{\psi}}{\partial\thr} }\ .  
 \end{array} 
 \right.
\label{system}
\end{eqnarray}
Straightforward integration of the last two equations gives 
\begin{eqnarray}
 \left\{ 
 \begin{array}{ll} 
 \displaystyle{A_{s}=-\frac{1}{L}\Phi_P(r)
   -\int_{0}^{r}\frac{\partial\widetilde{\psi}}{\partial s}
   \,{\rd}\bar r }\ ,\\[3mm]
 \displaystyle{A_\thr=\frac{1}{2\pi r}\Phi_T(r) 
   -\int_{0}^{r}\frac{\partial\widetilde{\psi}}{\partial\thr}\,{\rd}\bar r} \ .  
\end{array} \right.
\label{AS}
\end{eqnarray}
Let
\begin{equation}
\eta(r,\thr,s)=\int_{0}^r\widetilde{\psi}(\bar{r},\thr,s)\,{\rd}\bar{r}\ ,
\label{eta}
\end{equation}
also single-valued and periodic in $s$ and $\thr$; hence, eqs. (\ref{AS}) 
become 
\begin{eqnarray}
 \left\{ 
 \begin{array}{ll}
 \displaystyle{A_s=-\frac{1}{L}\Phi_P(r)
  -\frac{\partial\eta}{\partial s}+C_{1}(\thr,s)}\ ,\\[3mm]
 \displaystyle{A_\thr=\frac{1}{2\pi r}\Phi_T(r)
  -\frac{\partial\eta}{\partial\thr}+C_{3}(\thr,s)}\ , 
\end{array} 
\right.
\end{eqnarray}
where $C_{1}(\thr,s)$ and $C_{3}(\thr,s)$ are constants of 
integration: condition (\ref{condA}) is satisfied by setting 
$C_{1}(\thr,s)=\Phi_P(a)/L$, and continuity of $\bA$ at $r=0$ 
implies $C_3\left(\vartheta_R,s\right)=0$. Thus, we have
\begin{eqnarray}
 \left\{ 
 \begin{array}{ll}
 \displaystyle{A_s=\frac{\widetilde{\Phi}_P(r)}{L}}
  -\frac{\partial\eta}{\partial s}\ ,\\[3mm]
 \displaystyle{A_\thr=\frac{1}{2\pi r}\Phi_T(r)
  -\frac{\partial\eta}{\partial\thr}} \ , 
\end{array} 
\right.
\label{AZ}
\end{eqnarray}
where $\widetilde{\Phi}_P(r)=\Phi_P(a)-\Phi_P(r)$ is the complementary 
poloidal flux. Now let
\begin{equation}
H(r^{\ast})=\int_{V^{\ast}(K)}\bA\cdot\bB\,\rd^{3}\bx
  =\int_{0}^{L/k}\!\!\int_{0}^{2\pi}\!\int_{0}^{r^{\ast}}\!\!\!
  (A_\thr B_\thr+A_s B_s)\,{\rd}\bar r \, \bar r{\rd}\thr \, k{\rd} s\ ,
\label{subhelicity}
\end{equation}
the helicity associated with the tubular neighbourhood of radius 
$r^{\ast}$. As in CM95, by using the equations above 
and direct integration, we have
\begin{equation}
\displaystyle{H(r^{\ast})=\int_{0}^{r^{\ast}}\left(
  \frac{\rd\Phi_T}{{\rd}\bar r}\widetilde{\Phi}_{P}-\frac{\rd 
  \widetilde{\Phi}_P}{{\rd}\bar r}\Phi_T\right)\, {\rd}\bar r}\ ,
\label{hely}
\end{equation}
with total helicity given by
\begin{equation}
H=H(a)=\int_{0}^{a}\frac{\rd\Phi_T}{{\rd}\bar 
r}(\Phi_P(a)-\Phi_{P}(\bar r))\,{\rd}\bar r
  +\int_{0}^{a}\frac{\rd\Phi_P(\bar r)}{{\rd}\bar r}\Phi_T\, {\rd}\bar r\ ,
\label{hely2}  
\end{equation}
hence
\begin{equation}
H=2\int_{0}^{a}\Phi_T\frac{\rd\Phi_P}{{\rd}\bar r}\, {\rd}\bar r\ .
\label{hely3}
\end{equation}
By taking ${\rd}\Phi_P/{\rd} r=h({\rd}\Phi_T/{\rd} r)$, magnetic helicity 
is given by 
\begin{equation}
H=2h\int_{0}^{a}\Phi_T\frac{{\rd}\Phi_T}{{\rd} r}\,{\rd} r = h\Phi^2\ .
\label{helicitystandard}
\end{equation}
Hence, eq. (\ref{h}) reduces to eq. (\ref{hel}), with $h=Lk$.



\begin{thebibliography}{99}
\bibitem{Ar74}
  {\sc Arnold, V.I.} 1974
  The asymptotic Hopf invariant and its applications.
  In {\em Proc Summer School in Diff. Eqs. at Dilizhan},
  pp. 229--256. Armenian Acad. Sci. Erevan (in Russian). 
  [English translation: (1986) {\em Sel. Math. Sov. \bf5}, 
  327--345].
\bibitem{Bea04}
  {\sc Baranska, J., Pieranski, P., Przybyl, S. \& Rawdon, E.J.} 2004
  Length of the tightest trefoil knot. 
  {\em Phys. Rev. E \bf70}, 051810-1--051810-9.  
\bibitem{Bea05}
  {\sc Baranska, J., Pieranski, P. \& Rawdon, E.J.} 2005 
  Ropelength of tight polygonal knots.
  In {\em Physical and Numerical Models in Knot Theory} 
  (ed. J.A. Calvo \emph{et al.}), pp. 293--321.
  Series on Knots and Everything {\bf 36}, 
  World Scientific, Singapore. 
\bibitem{Ba78}
  {\sc Bateman, G.} 1978 {\em MHD instabilities}.  MIT Press.  
\bibitem{Be93}
  {\sc Berger, M.A.} 1993
  Energy--crossing number relations for braided magnetic fields.  
  {\em Phys. Rev. Lett. \bf70}, 705--708.  
\bibitem{BF84}
  {\sc Berger, M.A. \& Field, G.B.} 1984 
  The topological properties of magnetic helicity.  
  {\em J. Fluid Mech. \bf147}, 133--148.
\bibitem{BS99}
  {\sc Buck, G. \& Simon, J.} 1999 
  Thickness and crossing number of knots.  
  {\em Topol. Appl. \bf91}, 245--257. 
\bibitem{Ca61}
  {\sc C\u alug\u areanu, G.} 1961 
  Sur les classes d'isotopie des n\oe uds tridimensionnels et 
  leurs invariants. 
  {\em Czechoslovak Math. J.\/ \bf11}, 588--625.  
\bibitem{Cea02}
  {\sc Cantarella, J., Kusner, R.B. \& Sullivan, J.} 2002
  On the minimum ropelength of knots and links.
  {\em Inventiones Mathematicae\/ \bf150}, 257--286.
\bibitem{CM95}  
  [CM95] {\sc Chui, A.Y.K. \& Moffatt, H.K.} 1995
  The energy and helicity of knotted magnetic flux tubes.
  {\em Proc. Roy. Soc. Lond.\/ \rm  A \bf451}, 609--629.
\bibitem{Fr88}
  {\sc Freedman, M.H.} 1988
  A note on topology and magnetic energy in incompressible perfectly 
  conducting fluids. 
  {\em J. Fluid Mech. \bf194}, 549--551.  
\bibitem{FH91}
  {\sc Freedman, M.H. \& He, Z.-X.} 1991
  Divergence-free fields: energy and asymptotic crossing number. 
  {\em Ann. Math. \bf134}, 189--229.
\bibitem{Gd03}
  {\sc Gonzalez, O. \& De La Llave, R.} 2003
  Existence of ideal knots. 
  {\em J. Knot Theory \& Its Ramifications. \bf12}, 123--13 
\bibitem{GM90}
  {\sc Gonzalez, O. \& Maddocks, J.H.} 1990
  Global curvature, thickness, and the ideal shapes of knots.
  {\em Proc. Natl. Acad. Sci. USA \bf96}, 4769--4773. 
\bibitem{Lea99}
  {\sc Litherland, R.A., Simon, J., Durumeric, O. \& Rawdon, E.} 1999 
  Thickness of knots.  
  {\em Topol. Appl. \bf99}, 233--244.   
\bibitem{Me63}
  {\sc Mercier, C.} 1963 Sur une representation des surfaces
  toroidales: applications aux equilibres magnetohydrodynamiques.
  {\em Nucl.  Fusion\/ \bf3}, 89--98.  
\bibitem{Mo85}
  {\sc Moffatt, H.K.} 1985
  Magnetostatic equilibria and analogous Euler flows of arbitrarily
  complex topology. Part I. Fundamentals.
  {\em J. Fluid Mech. \bf159}, 359--378.  
\bibitem{Mo90}
  {\sc Moffatt, H.K.} 1990
  The energy spectrum of knots and links.  
  {\em Nature \bf347}, 367--369.
\bibitem{Mo01}
  {\sc Moffatt, H.K.} 2001
  Some remarks on topological fluid mechanics.  
  In {\em An Introduction to the Geometry and Topology of Fluid Flows}
  (ed. R.L. Ricca), pp. 3--10. NATO Science Series, II Mathematics, 
  Physics and Chemistry \textbf{47}. Kluwer Acad. Publs., Dordrecht, 
  The Netherlands.
\bibitem{MR92}
  {\sc Moffatt, H.K. \& Ricca, R.L.} 1992 
  Helicity and the C\u alug\u areanu invariant. 
  {\em Proc. R. Soc. Lond.\/ \rm A \bf439}, 411--429. 
\bibitem{Pi98} 
  {\sc Pieranski, P.} 1998
  In search of ideal knots.
  In {\em Ideal Knots} (ed. A. Stasiak \emph{et al.}), pp. 20--41. 
  Series on Knots and Everything {\bf 19}, 
  World Scientific, Singapore. 
\bibitem{PP02} 
  {\sc Pieranski, P. \& Przybyl, S.} 2002
  In search of the ideal trefoil knot.
  In {\em Physical Knots: Knotting, Linking, and Folding Geometric 
  Objects in $\RR^{3}$} (ed. J.A. Calvo \emph{et al.}), pp. 153--162. 
  Contemporary Mathematics {\bf 304}, 
  Am. Math. Soc., Providence (RI). 
\bibitem{Piea01} 
  {\sc Pieranski, P., Przybyl, S. \& Stasiak, A.} 2001
  Tight open knots.  
  {\em Eur. Phys. J.\/ \rm E \bf6}, 123--128.  
\bibitem{Pr01}
  {\sc Przybyl, S.} 2001
  The search for the ideal knots and their properties.
  {\em Ph.D. Thesis}. (In Polish). 
  Poznan University of Technology, Poznan. \newline
  \texttt{http://fizyka.phys.put.poznan.pl/\textasciitilde 
  pieransk/SylwesterPrzybyl-Thesis.pdf}.\newline
  [See also P. Pieranski's homepage: \newline
  \texttt{http://fizyka.phys.put.poznan.pl/\textasciitilde 
  pieransk/TablesUpTo9.html}.]  
\bibitem{Ri08}
  {\sc Ricca, R.L.} 2008 
  Topology bounds energy of knots and links. 
  {\em Proc. R. Soc. \rm A \bf464}, 293--300.   
\bibitem{Sp79}
  {\sc Spivak, M.} 1979
  {\em A comprehensive introduction to differential geometry}.
  Volume 1, Publish or Perish, Houston.
\bibitem{Sel98}      
  {\sc Stasiak, A., Katritch, V. \& Kauffman, L.H. (Eds.)} 1998
  {\em Ideal Knots.} 
  Series on Knots and Everything {\bf 19}, 
  World Scientific, Singapore.  
\bibitem{Wh69}      
  {\sc White, J.H.} 1969 
  Self-linking and the Gauss integral in higher dimensions. 
  {\em  Amer. J. Math.\/ \bf91}, 693--728.  
\bibitem{Wo58}
  {\sc Woltjer, L.} 1958
  A theorem on force-free magnetic fields.
  {\em Proc. Natl. Acad. Sci. USA \bf44}, 489--491.  
\end{thebibliography}
\end{document}